\documentclass[pra,preprint]{revtex4}%
\usepackage{amsfonts}
\usepackage{amsmath}
\usepackage{amssymb}
\usepackage{graphicx}%
\begin{document}

\title{Mechanocaloric and Thermomechanical Effects in Bose-Einstein Condensed Systems}

\author{G. C. Marques}
\altaffiliation[Corresponding author: ]{\texttt{marques@if.usp.br}}
\affiliation{Instituto de F\'{\i}sica, Universidade de S\~{a}o Paulo\\
Cx. Postal 66318, 05389-970 S\~{a}o Paulo, SP, Brazil}
\author{V. S. Bagnato}
\author{S. R. Muniz}
\altaffiliation[Current address: ]{School of Physics, Georgia Institute of Technology}
\affiliation{Instituto de F\'{\i}sica de S\~{a}o Carlos, Universidade de S\~{a}o Paulo\\
Cx. Postal 369, 13560-970 S\~{a}o Carlos, SP, Brazil}
\author{D. Spehler}
\affiliation{Universit\'{e} Louis Pasteur, I.U.T.\\
All\'{e}e d'Ath\`{e}nes, 67300 Schiltigheim, France}

\begin{abstract}
In this paper we extend previous hydrodynamic equations, governing the motion of Bose-Einstein-condensed fluids, to include temperature effects. This allows us to analyze some differences between a normal fluid and a Bose-Einstein-condensed one. We show that, in close analogy with superfluid He-4, a Bose-Einstein-condensed fluid exhibits the mechanocaloric and thermomechanical effects. In our approach we can explain both effects without using the hypothesis that the Bose-Einstein-condensed fluid has zero entropy. Such ideas could be investigated in existing experiments.

\end{abstract}
\maketitle

\section{\noindent\textbf{INTRODUCTION} \vspace{0.3cm}}

The achievement of Bose-Einstein condensation (BEC) in dilute atomic gases
\cite{1} represents the establishment of several new exciting possibilities.
The investigation of degenerate quantum gases provides, for instance, a good
testing ground for innumerable many-body theories developed along the past few
decades. Those theories were mainly developed within the context of explaining
the striking properties of Helium-4 below the $\lambda$-point. Among other
features, this fluid behaves as having zero viscosity (a superfluid) and
apparently also as having zero entropy \cite{2}.

The zero entropy hypothesis seemed to be the only explanation for two other
(although related) remarkable properties of superfluid He-4. The first notable
property is the called thermomechanical effect (or the \textquotedblleft
fountain\textquotedblright\ effect), where a temperature gradient across the
fluid produces motion of matter. The force responsible for imparting the
motion to the fluid is a \textquotedblleft thermoforce\textquotedblright\ (a
term used first by London \cite{3}). On the other hand, the motion of the
superfluid also produces temperature gradients. This second effect is the so
called mechanocaloric effect. Both effects were very important to understand
the behavior of superfluids, giving valuable hints about the nature of the
processes involved.

In earlier times the first proposal of London and Tisza \cite{3,Tisza},
considering the superfluidity as a macroscopic manifestation of BEC, was
somehow overruled by the good agreement of the hydrodynamic theory of Landau
\cite{Landau} to the first experiments measuring the second sound velocity
\cite{Pellam}. Just later, after much of the development of the many-body
theory, specially by Bogoliubov \cite{Bogoliubov}, it became clear that BEC
was in fact behind the appearance of superfluidity. Even then, the
hydrodynamic formalism has been quite useful for calculating and understanding
the He-4 superfluid. Recently, after the experimental achievement of quantum
degenerate gases, the formalisms developed initially to understand
superfluidity of Helium have been extended to explore the dilute atomic gases
as well \cite{Pethick}. A full historical account of these developments,
although exciting, is much beyond the scope of this present paper. The
interested reader, however, could look for instance at reference
\cite{Griffin} and the references therein.

In this paper we deal with some aspects of the hydrodynamics of Bose-Einstein
condensed phase. In particular we show that both effects (the mechanocaloric
and the thermomechanical effects) are always expected to happen in any
Bose-Einstein condensate.

The way we deal with these two phenomena are derived from the same equation,
showing the clear relation between them. Furthermore, we do not have to use
the zero-entropy property for the superfluid or the BEC system to explain both effects.

The presentation of this paper is arranged as follows. In section~II we
introduce the hydrodynamic quantization approach to the Bose-Einstein
condensation of a system of charged spin-0 bosons under the action of external fields. A
chemical potential is introduced later in order to discuss the equilibrium
conditions for this system. In section~III we present the equations of motion
and define averages over the ensemble.

In section IV we write the hydrodynamic equations of Bose-Einstein condensates
under the action of external fields. These equations do not take into account
temperature effects.

The relevant equilibrium equation, a generalized Bernoulli equation, is
introduced in section~V. From this zero temperature equilibrium condition we
then propose a natural extension that allows us to take into account also the
temperature effects.

The force equation and the explicit expression for the thermoforce is
presented in section~VI. Finally, the conclusions in section~VII close the paper.

\bigskip

\section{\textbf{HYDRODYNAMIC QUANTIZATION} \vspace{0.3cm}}

In earlier papers \cite{4,5} we have presented the hydrodynamic quantization
approach and some aspects of the fluidity of Bose-Einstein condensed system.
In this section we shall analyze these equations for the quite general and
important case of a condensate in the presence of electric and magnetic fields
and when the number of particles varies. In order to do so we introduce a
chemical potential. Furthermore, the introduction of a chemical potential is
also relevant in order to study the equilibrium conditions for condensation
under the action of external fields \cite{6}.

We shall discuss here the field theoretic approach aimed at the description of
a system of scalar Bose particles. Within the field theoretic approach, charged
spin 0 bosons are associated to a complex scalar field $\,\psi(x)$. In what
follows, $\,x\,$ is a short hand notation for a time and space point
($x=(t,\vec{x}\,)$).

The quantization of the theory is carried out by imposing equal-time
commutation relations between the field and its canonically conjugated
variable $\,\psi^{\ast}$, followed by a typical quantization scheme
implemented with the introduction of the Fock space. This method is especially
useful when we deal with systems whose number of particles is well defined,
for example in the description of scattering processes. One can use, however,
an alternative procedure: the so-called hydrodynamic quantization of the
field. In this scheme we use a new set of variables $\,\rho(x)\,$ and
$\,\varphi(x)$, which are more convenient in the description of properties
related to Bose-Einstein condensation. These new variables are defined as:

\begin{equation}%
\begin{array}
[c]{l}%
\rho(x)\,=\,\psi^{\ast}(x)\,\psi(x)\\[0.2cm]%
\psi(x)\,=\,\sqrt{\rho(x)}\;e^{i\varphi(x)}\ ,
\end{array}
\tag*{(II.1)}%
\end{equation}
denominated density and phase variables, respectively. They are canonically
conjugated to each other, as we shall see below. For this reason we impose the
following commutation relations among the density operator and the phase operator%

\begin{equation}
\left[  \,\widehat{\rho}(x)\,,\,\widehat{\varphi}(x^{\prime})\right]
\,=\,i\delta(x-x^{\prime})\ . \tag*{(II.2)}%
\end{equation}

The quantization method, based on commutation relations among density and
phase operators, is called a hydrodynamic quantization, and its application in
superfluidity, where Bose-Einstein condensation is known to occur, was
proposed by Landau \cite{6}. The commutation relation in equation (II.2), in
conjunction with an explicit representation for these operators, requires a
departure from more usual quantization approaches.

An explicit representation of the algebra of operators (II.2) can be realized
by making use of the so-called density representation. In this representation
the density operator is a classical $c$-number. That is,
\begin{equation}
\widehat{\rho}\,(x)=\rho(x) \tag*{(II.3)}%
\end{equation}
and the phase operator is represented by the operator
\begin{equation}
\widehat{\varphi}(x)\,=\,-\,i\,\frac{\delta}{\delta\rho(x)}\ \ . \tag*{(II.4)}%
\end{equation}

Within the density representation the states vectors are represented as
functionals of the density
\begin{equation}
\psi=\psi\lbrack\rho]\ . \tag*{(II.5)}%
\end{equation}

In Ref.\cite{4} we give examples of how to construct wave functionals
associated to the vacuum and Fock states.

The state functionals, as well as other relevant physical quantities, are
written in terms of the density and phase operators (II.1), or in terms of
integrals over space-time densities. We shall give below several examples that
will be used through this paper. We start with the Lagrangian and the Hamiltonian.

In terms of the field $\,\psi$, the classical Lagrangian density
$\,\mathcal{L}$, for the nonrelativistic scalar particles, is written as
\vspace{0.2cm}
\begin{equation}
\mathcal{L}\,=\,\frac{i}{2}\,\left\{  \psi^{\ast}(x)\left[  \partial_{t}%
\,\psi(x)\right]  -\left[  \partial_{t}\,\psi^{\ast}(x)\right]  \psi
(x)\right\}  -\frac{\vec{\nabla}\psi^{\ast}(x)\cdot\vec{\nabla}\psi(x)}%
{2m}-\mathcal{H}^{I}\left[  \psi^{\ast}(x)\,\psi(x)\right]  \ , \tag*{(II.6)}%
\end{equation}

\vspace{0.2cm}

\noindent where $\,\mathcal{H}^{I}\,$ is the interaction Hamiltonian density.
We define the Hamiltonian density $\,\mathcal{H}\,$ as composed of two terms:
\vspace{0.2cm}
\begin{equation}
\mathcal{H}(x)\;\equiv\;\frac{\vec{\nabla}\psi^{\ast}(x)\cdot\vec{\nabla}%
\psi(x)}{2m}+\mathcal{H}^{I}\left[  \psi^{\ast}(x)\,\psi(x)\right]
\;\equiv\;K+\mathcal{H}^{I}\ . \tag*{(II.7)}%
\end{equation}

\vspace{0.2cm}

The first term in (II.7), containing derivatives of the field, is the kinetic
energy term ($K$) and the second is the interaction Hamiltonian $\,\mathcal{H}%
^{I}$, which contains no field derivative terms.

Placing equation~(II.1) into equation~(II.6) one can see that the classical
Lagrangian density can be written under the general form: \vspace{0.2cm}
\begin{equation}
\mathcal{L}(x)\,=\,-\,\rho(x)\,\frac{\partial\varphi(x)}{\partial
t}-\mathcal{H}[x,\rho,\varphi]\ . \tag*{(II.8)}%
\end{equation}
Thus showing that phase and density are canonically conjugated variables.

The quantum action is defined as
\begin{equation}
S=\int\mathcal{L}(x)\,dx\ . \tag*{(II.9)}%
\end{equation}

If the system is under the action of external magnetic fields ($\vec{B}%
=\vec{\nabla}\times\vec{A}\,$) the kinetic term can be written, in terms of
the variables $\,\rho(x)\,$ and $\,\varphi(x)$, as: \vspace{0.2cm}
\begin{equation}
K\left(  \rho\,,\,\varphi\,,\,\vec{A}\,\right)  =\rho(x)\,\frac{\left[
\vec{\nabla}\varphi(x)-e\vec{A}(x)\right]  ^{2}}{2m}+\frac{\left(  \vec
{\nabla}\,\sqrt{\rho(x)}\,\right)  ^{2}}{2m} \tag*{(II.10)}%
\end{equation}

\vspace{0.2cm}

\noindent where $\, e\,$ is the electric charge of the spin~0 bosons.

Whereas, in the presence of an external potential $\,U(x)$, the interaction
Hamiltonian takes the general form:
\begin{equation}
\mathcal{H}^{I}(\rho)=\rho(x)\,U(x)+\rho(x)\,\varepsilon(\rho,x)\;\equiv
\;\rho(x)\,U(x)+\mathcal{H}^{\mathrm{int}} \tag*{(II.11)}%
\end{equation}
where $\,\varepsilon(\rho,x)\,$ in (II.11) is the per-particle interaction
internal energy density. If we assume binary interactions among the particles,
$\,\varepsilon(\rho,x)\,$ is given by \vspace{0.2cm}
\begin{equation}
\varepsilon(\rho,x)\,=\,\frac{1}{2}\int d\vec{x}\,^{\prime}\,V(\vec{x}-\vec
{x}\,^{\prime})\,\rho(\vec{x}^{\prime})\ . \tag*{(II.12)}%
\end{equation}

Let us consider the expression, in terms of density and phase, of other
physically relevant quantities in the study of quantum fluids. We start with
the expression for the momentum. The classical definition of momentum is the
integral over the space of the momentum operator density ($\vec{\mathcal{P}%
}(x)$) of the field: \vspace{0.2cm}
\begin{equation}
\vec{p}=\int d\vec{x}\,\vec{\mathcal{P}}(\vec{x}\,)=\,\frac{1}{2}\int
dx\,\varphi^{\ast}(x)\,\overleftrightarrow{\vec{\nabla}}\,\varphi(x)=\int
d^{3}\vec{x}\,\rho(x)\,\vec{\nabla}\varphi(x)\ . \tag*{(II.13)}%
\end{equation}

\vspace{0.2cm}

From (II.13) it follows that the per-particle momentum $\,\vec{P}(\vec
{x}\,)\,$ is given by \vspace{0.2cm}
\begin{equation}
\vec{P}(\vec{x})=\vec{\nabla}\varphi(x)=\vec{\nabla}\left(  -\,i\,\frac
{\delta}{\delta\rho(x)}\right)  \equiv\,m\vec{V}(\vec{x}\,)\ . \tag*{(II.14)}%
\end{equation}

\vspace{0.2cm}

In quantum theory there is, however, another relevant momentum. This quantum
momentum ($\vec{P}_{q}(x)$) is defined as $\dfrac{1}{2}\,$ of the gradient of
the log of the density. That is
\begin{equation}
\vec{P}_{q}(\vec{x}\,)\;\equiv\;\vec{\nabla}\left(  \ln\,\sqrt{\rho
(x)}\,\right)  =m\,\vec{V}_{q}(x)\ . \tag*{(II.15)}%
\end{equation}
The quantum origin of $\,\vec{P}_{q}(\vec{x}\,)\,$ defined in (II.15) is
explained in Ref. \cite{8}.

The analysis of a hard sphere gas provides a good example for the
understanding of both types of momenta \cite{5}. The classical momentum
$\,\vec{P}(\vec{x}\,)\,$ is associated to the quantized vortices, whereas the
quantum momentum $\,\vec{P}_{q}\,$ is more relevant in the understanding the
dependence of the fluid density upon the distance of the vortex center
\cite{5}.

In terms of the velocities defined in (II.13) and (II.14) one can write, by
using (II.10) and (II.11), the Hamiltonian density $\,\mathcal{H}\,$ as
\vspace{0.2cm}
\begin{equation}
\mathcal{H}[\rho,\varphi]=\rho(x)\,\frac{m}{2}\left[  \vec{V}^{2}(x)+\vec
{V}_{q}^{2}(x)\right]  +\rho(x)\big[U(x)+e(\rho,x)\big]\ . \tag*{(II.16)}%
\end{equation}

\vspace{0.2cm}

If the bosonic system is under the action of an external magnetic field the
above expression remains valid for $\,\vec{V}(x)\,$ defined by \vspace{0.2cm}
\begin{equation}
\vec{V}(x)\,=\,\frac{\vec{P}(x)-e\vec{A}(x)}{m}\ \ . \tag*{(II.17)}%
\end{equation}

The interaction of the particles with external fields is carried out, in field
theory, as usual. For the coupling with external magnetic fields ($\,\vec B =
\vec\nabla\times\vec A\,$) we use the minimum substitution $\,\vec
\nabla\rightarrow\vec A - ie\vec A\,$. The potential $\, U(x)\,$ takes into
account the interaction with external electric fields as well as external
gravitational fields.

All the above definitions are relevant in the understanding of fluidity
aspects of Bose-Einstein condensed systems.

In the following section we shall analyze the dynamical equations.

\bigskip

\section{\textbf{EQUATIONS OF MOTION AND ENSEMBLE AVERAGES} \vspace{0.3cm}}

The time evolution of a physical quantity represented by the operator
$\,O(x)\,$ is given by the Heisenberg equation of motion,
\begin{equation}
\frac{\partial O(x)}{\partial t}\,=\,i\left[  H,O(x)\right]  \ .
\tag*{(III.1)}%
\end{equation}

The Hamiltonian operator in (III.1) is the integral of the Hamiltonian
density
\begin{equation}
H=\int dx\,\mathcal{H}[\varphi,\rho]\ . \tag*{(III.2)}%
\end{equation}

Within the density representation the equations of motion of the density and
phase operators are:
\begin{subequations}
\begin{align}
&  \frac{\partial\varphi(x)}{\partial t}\,=\,-\,\frac{\delta H}{\delta\rho
(x)}\ ,\tag*{(III.3)}\\[0.12in]
&  \frac{\partial\,\rho(x)}{\partial t}\,=\,\frac{\delta H}{\delta\varphi
(x)}\ . \tag*{(III.4)}%
\end{align}

\vspace{0.2cm}

These equations follow also from (II.8) by taking $\, \rho\,$ and $\,
\varphi\,$ as independent dynamical variables.

By taking the gradient of (III.3) we define an yet new equation that we name
the force equation:
\end{subequations}
\begin{equation}
\frac{\partial\vec P(x)}{\partial t} \, = \, \vec F(x) \ . \tag*{(III.5)}%
\end{equation}

\vspace{0.2cm}

\noindent The force equation gives the rate of change of the per-particle
momentum and it is the relevant equation in the understanding of superfluidity
and the thermoforce. As we shall see, the thermoforce is a new type of force
that arises in a BEC system.

The local force operator $\,{\vec{F}}(x)\,$ is, formally, written as
\begin{equation}
\vec{F}(x)\;=\;-\,\vec{\nabla}\left(  \frac{\delta H}{\delta\rho(x)}\right)
\,. \tag*{(III.6)}%
\end{equation}

One of the interesting features of the hydrodynamic quantization is that the
quantum equation of motion resembles that of a classical fluid (from this fact
derives the \textquotedblleft hydrodynamic\textquotedblright\ name). Using
Hamiltonian (II.16), the time evolution equations are \vspace{0.2cm}
\begin{align}
&  -\,\frac{\partial\varphi(x)}{\partial t}\,=\,\frac{\vec{P}^{2}(x)}%
{2m}+U(x)+h(x)+\frac{\vec{P}_{q}^{2}}{2m}+\frac{1}{2m\rho}\,\vec{\nabla}%
\cdot\left(  \rho\,\vec{P}_{q}\right)  \ ,\tag*{(III.7)}\\[0.12in]
&  \quad\frac{\partial\rho\,(x)}{\partial t}\,=\,-\vec{\nabla}\,\cdot
\,\frac{\rho(x)\,\vec{P}(x)}{m}\ ,\tag*{(III.8)}\\[0.3cm]
&  \quad\frac{\partial\vec{P}}{\partial t}\,=\,-\vec{\nabla}\left(  \frac
{\vec{P}^{2}}{2m}\right)  +\vec{F}_{\mathrm{ext}}(x)-\vec{\nabla}%
h(x)-\vec{\nabla}\left(  \frac{\vec{P}_{q}^{2}}{2m}+\frac{1}{2m\rho}%
\,\vec{\nabla}\cdot\left(  \rho\,\vec{P}_{q}\right)  \right)  \,.\qquad
\quad{\ } \tag*{(III.9)}%
\end{align}

\vspace{0.2cm}

Where $\,h(x)\,$ in (III.7) and (III.9) is the per-particle enthalpy
\vspace{0.2cm}
\begin{equation}
h(x)\;\equiv\;\frac{\delta\,\mathcal{H}^{\mathrm{int}}}{\delta\rho
(x)}\,=e(\rho,x)+\frac{P(x)}{\rho(x)}\ . \tag*{(III.10)}%
\end{equation}

\vspace{0.2cm}

The pressure $\,P(x)\,$ in (III.10) is, as can be inferred from (III.10) and
(II.11), given by \vspace{0.2cm}
\begin{equation}
\frac{P(x)}{\rho(x)}\,=\int dx^{\prime}\,\rho(x^{\prime})\,\frac{\delta
e(\rho,x^{\prime})}{\delta\rho(x)}\,e\left(  \rho(x^{\prime}),x^{\prime
}\right)  \ . \tag*{(III.11)}%
\end{equation}

\vspace{0.2cm}

The term $\,\vec{F}_{\mathrm{ext}}(x)\,$ in equation(III.9) is the force
exerted on each of the charged particles of the system as a result of the
external potential $\,U(x)$,
\begin{equation}
\vec{F}_{\mathrm{ext}}(x)\,=\,-\vec{\nabla}\,U(x)\ . \tag*{(III.12)}%
\end{equation}

Equations (III.7)--(III.9) are valid as time evolution equations of operators.
In order to deal with classical $c$-numbers one can consider expectation
values and averages over the ensemble.

Any of the quantum versions of the equations of motion (III.7)--(III.9), are
obviously valid when one considers expectation values. Defining the trace of
an operator $\,O\,$ as
\begin{equation}
T_{r}\,O(x)=\sum_{\psi}\left\langle \psi\left\vert O(x)\right\vert
\psi\right\rangle \,\frac{1}{\left\langle \psi|\psi\right\rangle }\ ,
\tag*{(III.13)}%
\end{equation}
we can write the following equations: \vspace{0.2cm}
\begin{align}
&  \frac{\partial}{\partial t}\,\left(  T_{r}\,\varphi(x)\right)
\,=\,-\,T_{r}\left(  \frac{\delta\mathcal{H}}{\delta\rho(x)}\right)
\,,\tag*{(III.14)}\\[0.16in]
&  \frac{\partial}{\partial t}\,\left(  T_{r}\,\rho(x)\right)  \;=\;T_{r}%
\left(  \frac{\delta\mathcal{H}}{\delta\varphi(x)}\right)  \,. \tag*{(III.15)}%
\end{align}

\vspace{0.2cm}

Ehrenfest theorems can be defined as averages over the ensemble of quantum
equations. In order to define averages over the ensemble, we define the
quantum-mechanical partition function of a Bose system.

For the quantum action $\,S$, given by (II.9), we define the partition
function as
\begin{equation}
Z=T_{r}\;e^{iS}\ . \tag*{(III.16)}%
\end{equation}

This is just an extension of the usual finite temperature ($T$) definition of
the partition function:
\begin{equation}
Z=T_{r}\;\left(  e^{-\frac{H}{kT}}\right)  \ . \tag*{(III.17)}%
\end{equation}

The average over the ensemble of a physical quantity $\,O(x)\,$ is usually
defined by
\begin{equation}
\langle O(x)\rangle\,=\,\frac{1}{Z}\,T_{r}\left\{  e^{iS}\,O(x)\right\}  \ .
\tag*{(III.18)}%
\end{equation}

Taking averages over the ensemble, one writes: \vspace{0.2cm}
\begin{align}
\frac{\partial\langle\varphi(x)\rangle}{\partial t}\,  &  =\,-\,\left\langle
\frac{\delta\mathcal{H}}{\delta\rho(x)}\right\rangle \ , \tag*{(III.19)}%
\\[0.16in]
\frac{\partial\langle\rho(x)\rangle}{\partial t}\,  &  =\,+\,\left\langle
\frac{\delta\mathcal{H}}{\delta\varphi(x)}\right\rangle \ . \tag*{(III.20)}%
\end{align}

Averages over the ensemble are relevant in the context of statistical
mechanics. In the next section we shall consider matrix elements of the
equations of motion under the form (III.14) and (III.15).

\bigskip

\section{HYDRODYNAMIC EQUATIONS FOR BOSE-EINSTEIN CONDENSED STATES}

One expects some basic distinctions, at the level of states, between a
condensed system and a normal one. We have proposed, in Ref. \cite{4}, that
the basic distinction can be traced back to special properties of the wave
functional associated to BEC states. We have proposed that the wave functional
of BEC systems is endowed with two distinctive properties \cite{4,5}:

\bigskip

\textbf{Property (1) of the condensate wave functional.}

The wave function of the system associated to $\,\psi_{c}[\rho]\,$ \cite{5,9}
is a product of wave functions
\begin{equation}
\psi_{c}\left(  x_{1}\ldots x_{N}\right)  =\prod_{i=1}^{N}\psi_{c}\left(
x_{i}\right)  \ . \tag*{(IV.1)}%
\end{equation}
The wave function $\,\psi_{c}(x)\,$ in (III.1) is, by definition, the
condensate wave function which will be written as
\begin{equation}
\psi_{c}(x)\,=\,\sqrt{\rho_{c}(x)}\;e^{i\varphi_{c}(x)}\ . \tag*{(IV.2)}%
\end{equation}

The variables $\, \rho_{c}(x)\, $ and $\, \varphi_{c}(x)\,$ will be identified
as the density and phase of the condensate.

\bigskip

\textbf{Property (2) of the condensate wave functional.}

The condensate wave functional describes coherent states for which the
following factorization property holds true: \vspace{0.2cm}
\begin{equation}
\dfrac{\psi^{\ast}[\rho]\,\widehat{\psi}(x)\,\psi\lbrack\rho]}{\psi^{\ast
}[\rho]\,\psi\lbrack\rho]}\,=\psi_{c}(x)=\,\sqrt{\rho_{c}(x)}\;e^{i\varphi
_{c}(x)} \tag*{(IV.3)}%
\end{equation}
and
\begin{equation}
\dfrac{\psi^{\ast}[\rho]\cdot\widehat{\psi}(x_{1})\,\ldots\,\widehat{\psi
}(x_{N})\,\psi\lbrack\rho]}{\psi^{\ast}[\rho]\,\psi\lbrack\rho]}\,=\psi
_{c}(x_{1})\,\ldots\,\psi_{c}(x_{N})\ . \tag*{(IV.4)}%
\end{equation}

\vspace{0.2cm}

\bigskip

Property~(1) is an obvious requirement in order that the wave functional be
associated to a condensed state. Whereas property~(2) allows us to identify
the condensate wave function with the order parameter of the phase
transition \cite{10,11}.

The factorization property (III.4) is the analog of the Off Diagonal Long
Range Order (ODLRO) introduced by Penrose and Onsager \cite{12} in the
context of superfluidity of He-4.

In References \cite{4} and \cite{5} we have shown that the wave
functional
\begin{equation}
\psi_{c}[\rho]=e^{\int\rho(x)\ln\psi_{c}(x)dx} \tag*{(IV.5)}%
\end{equation}
exhibits the properties above mentioned properties (1) and (2). Thus providing
an explicit example of such wave functionals.

The relevance of the factorization property in the understanding of
superfluidity has been already emphasized by Anderson in Ref. \cite{10}. By
taking averages over states satisfying (IV.4), we conclude that the wave
function of the condensate, defined in (IV.3), satisfy the equations:
\vspace{0.2cm}
\begin{align}
\frac{\partial\varphi_{c}(x)}{\partial t}  &  \!\!=\!\!\left.  -\left(
\frac{\delta\mathcal{H}}{\delta\rho(x)}\right)  \right\vert _{\overset
{{\scriptstyle\rho=\rho_{c}(x)}}{{\scriptstyle\varphi=\varphi_{c}(x)}}%
}\tag*{(IV.6)}\\[0.2in]
\frac{\partial\rho_{c}(x)}{\partial t}  &  \!\!=\!\!\left.  \left(
\frac{\delta\mathcal{H}}{\delta\varphi(x)}\right)  \right\vert _{\overset
{{\scriptstyle\rho=\rho_{c}(x)}}{{\scriptstyle\varphi=\varphi_{c}(x)}}}\ \ .
\tag*{(IV.7)}%
\end{align}

\vspace{0.2cm}

Since a $c$-number field $\, \psi(x)\,$ is the wave function in the $\vec
x$-representation of a state $\, |\psi\rangle\,$, equations (IV.6) and (IV.7)
specify the equation for the condensate wave function. They are $c$-number equations.

By taking the functional derivatives of $\,\mathcal{H}\,$ given by (II.2), now
expressed as a functional of $\,\rho\,$ and $\,\varphi$, we obtain the
following set of equations: \vspace{0.2cm}
\begin{align}
&  \quad\frac{\partial\rho_{c}(x)}{\partial t}+\vec{\nabla}\cdot\vec{J}%
_{c}(x)=0\tag*{(IV.8)}\\[0.12in]
&  -\,\frac{\partial\varphi_{c}(x)}{\partial t}\,=\,\frac{m}{2}\,\vec{V}%
^{2}(x)+\frac{m}{2}\,\vec{V}_{q}^{2}(x)-\frac{\vec{\nabla}\cdot\vec{J}_{q}%
(x)}{2\rho_{c}(x)}+U(x)+h(x)\ .\qquad\quad{\ } \tag*{(IV.9)}%
\end{align}

\vspace{0.2cm}

\noindent Where we have used for $\,\vec{V},\ \vec{V}_{q}\,$ and $\,h(x)\,$
the definition (II.7), (II.15) and (III.10). The density currents $\,\vec
{J}_{c}\,$ and $\,\vec{J}_{q}\,$ are given by:
\begin{align}
\vec{J}_{q}(x)  &  \!\!=\!\!\rho_{c}(x)\,\vec{V}_{q}(x) \tag*{(IV.10)}%
\\[0.08in]
\vec{J}_{c}(x)  &  \!\!\equiv\!\!\rho_{c}(x)\,\vec{V}_{c}(x)\ . \tag*{(IV.11)}%
\end{align}

Taking the gradient of (III.9) we get the force equation: \vspace{0.2cm}
\begin{equation}
\frac{\partial\vec{P}_{c}}{\partial t}+m\left(  \vec{V}_{c}\cdot\vec{\nabla
}\right)  \vec{V}_{c}+m\left(  \vec{V}_{q}\cdot\vec{\nabla}\right)  \vec
{V}_{q}=e\,\vec{V}_{c}\wedge\vec{B}-\vec{\nabla}\left\{  U+h+\,\frac{1}{2\rho
}\,\vec{\nabla}\cdot\rho\,\vec{V}_{q}\right\}  \ . \tag*{(IV.12)}%
\end{equation}

In the presence of an external magnetic field the motion is necessarily
rotational since the rotational of the condensate velocity is related to the
magnetic field by
\begin{equation}
\vec{\nabla}\times\vec{V}_{c}\,=\,-\,\frac{e}{m}\,\vec{\nabla}\times\vec
{A}\,=\,-\,\frac{e}{m}\,\vec{B}\ . \tag*{(IV.13)}%
\end{equation}

\vspace{0.2cm}

In the absence of a magnetic field the motion of the condensate is irrotational.

Equation (III.8) is continuity equation which we have shown to be true for the
condensed component of the system.

As we shall see in the next section, equation (III.9) corresponds to a
generalized Bernoulli equation. We refer to it as the equilibrium equation. It
gives the condition of equilibrium for a Bose-Einstein condensed fluid under
the action of external fields.

Equation (III.13) is the dynamic equation satisfied by the fluid in motion
under external fields. We shall see that it is an extension of one of
Anderson's equations in Ref.~\cite{10}.

This allows us to conclude that the dynamical equations for the density and
phase of the wave function of the condensate, are (IV.8), (IV.9) and (IV.12).
As a result of these equations we can conclude, on quite general grounds, that
the condensed fluid satisfy continuity equation and that in the absence of
magnetic fields the motion of the fluid is irrotational.

\bigskip

\section{\textbf{EQUILIBRIUM EQUATION} \vspace{0.3cm}}

We show in this section that, for stationary states, equation (IV.9) gives the
equilibrium condition when the system is under the action of external fields
and that this equilibrium condition is a generalized Bernoulli equation.

As usual, we assume that the equilibrium condition for a system under the
action of external fields is the chemical potential ($\mu$) to be constant:
\begin{equation}
\mu=\mu_{0}\ . \tag*{(V.1)}%
\end{equation}

Remembering that the field $\,\psi\,$ is the wave-function of a state
$\,\psi\,$ in the $\vec{r}$-representation, the field associated to a
stationary state is
\begin{equation}
\psi(\vec{x},t)=e^{i\big(\mu_{0}t+\varphi_{c}(\vec{x}\,)\big)}\;\sqrt{\rho
_{c}(\vec{x}\,)} \tag*{(V.2)}%
\end{equation}
where $\,\mu_{0}\,$ is a constant.

For a stationary state the equilibrium equation, for the phase given by (V.2),
is
\begin{equation}
\mu_{0}=h(x)+U(x)+\frac{m\,\vec{V}_{c}^{2}(x)}{2}+\frac{m}{2}\,\vec{V}_{q}%
^{2}(x)-\,\frac{\vec{\nabla}\cdot J_{q}(x)}{2\rho_{c}(x)}\ \ . \tag*{(V.3)}%
\end{equation}
Taking now into account the explicit expression for $\,h(x)$, we write
\vspace{0.2cm}
\begin{equation}
\mu_{0}=\varepsilon\big(x,\rho_{c}(x)\big)+\frac{P(x)}{\rho_{c}(x)}+\frac
{m}{2}\,\vec{V}_{c}^{2}(x)+U(x)+\frac{m}{2}\,\vec{V}_{q}^{2}(x)-\frac
{\vec{\nabla}\cdot\vec{J}_{q}(x)}{2\rho_{c}(x)}\ . \tag*{(V.4)}%
\end{equation}

\vspace{0.2cm}

In order to show that $\,\mu_{0}\,$ is the chemical potential we multiply
equation (V.4) by $\,\rho_{c}(x)\,$ and integrate over $\,\vec{x}\,$. The
result can be written as
\begin{equation}
\mu_{0}\,N_{c}\,=\,E_{c}+\int P(\vec{x}\,)\,d^{3}\vec{x} \tag*{(V.5)}%
\end{equation}
where $\,N_{c}\,$ in (V.5) is the number of particles in the condensate
\begin{equation}
N_{c}=\int d^{3}\vec{x}\,\rho_{c}(\vec{x}\,) \tag*{(V.6)}%
\end{equation}
and $\,E_{c}\,$ is the condensate energy defined as the integral of the
density Hamiltonian:
\begin{equation}
E_{c}=\int d^{3}\vec{x}\,\mathcal{H}\big[\rho_{c},\varphi_{c}\big]\ .
\tag*{(V.7)}%
\end{equation}

Expression (V.5) shows that $\,\mu_{0}\,$ is the chemical potential since
$\,\mu_{0}N_{c}\,$ is the total enthalpy of the system \cite{6}.

For a uniform condensate, the Bogoliubov condensate \cite{14},
\begin{equation}
\rho_{c}(x)=\rho_{0c}\ . \tag*{(V.8)}%
\end{equation}
The quantum velocity vanishes so that for a uniform condensate,
equation~(IV.3) is Bernoulli's equation: \vspace{0.2cm}
\begin{equation}
\mu_{0}=\varepsilon\big(x,\rho_{0c}(x)\big)+m\,\frac{P(x)}{\rho_{0c}}+\frac
{m}{2}\,\vec{V}_{c}^{2}(x)+U(x)\ . \tag*{(V.9)}%
\end{equation}

\vspace{0.2cm}

Since the first four terms of equation (IV.4) corresponds to the usual
Bernoulli equation, one can predict that deviations from the usual fluid is
expected as a result of the quantum velocity term. This prediction deserves to
be tested experimentally.

We notice that for a non self interacting system
\begin{equation}
h(x)=0\ . \tag*{(V.10)}%
\end{equation}
Such non self interacting fluid, the fluid described by (IV.2) and (IV.3), is
a Mahdelung fluid \cite{13}. Therefore, neglecting self interaction of
particles, the BEC is a Mahdelung fluid.

One can extend some of the previous equations to finite temperatures. In order
to do so we just recall that the equilibrium condition for a fluid under
external fields takes the form \cite{6}:
\begin{equation}
\mu=g(P,T)+\mu^{\prime} \tag*{(V.11)}%
\end{equation}
where $\,g(P,T)\,$ is the per-particle Gibbs energy in the absence of the
external fields and $\,\mu^{\prime}\,$ takes into account the external fields.

Remembering also that at zero temperatures the per-particle Gibbs free energy
is equal to the per-particle enthalpy
\begin{equation}
g(P,0)=\varepsilon\left(  x,\rho_{0c}\right)  +m\,\frac{P(x)}{\rho_{0c}%
(x)}\,=h(x) \tag*{(V.12)}%
\end{equation}
we can write equation (IV.6) under the following form
\begin{equation}
\mu_{0}=g(P,0)+\,\frac{m}{2}\,\vec{V}_{c}^{2}(x)+U(x)\ . \tag*{(V.13)}%
\end{equation}

The natural extension of (V.5) to finite temperatures, in view of (V.13), is
\vspace{0.2cm}
\begin{align}
\mu_{0}  &  \!\!=\!\!g(P,T)+\frac{m\,\vec{V}_{c}^{2}(x)}{2}%
+U(x)\nonumber\\[0.12in]
&  \!\!=\!\!\varepsilon\left(  x,\rho_{0}\right)  +m\,\frac{P(x)}{\rho_{0}%
}-Ts(T)+m\,\frac{\vec{V}_{c}^{2}}{2}+U(x)\ . \tag*{(V.14)}%
\end{align}

\vspace{0.2cm}

Notice that for non stationary states the generalization of (IV.11) to finite
temperatures should be, from (V.2) and (V.14) \vspace{0.2cm}
\begin{equation}
\frac{\partial\varphi_{c}}{\partial t}\,=\,g(P,T)+m\,\frac{\vec{V}_{c}^{2}}%
{2}\,+U(x)\ . \tag*{(V.15)}%
\end{equation}

\vspace{0.2cm}

The interesting aspect of (V.14) is that one can imagine an \emph{Entropy
Filter\/}. That is, one can devise experiments by means of which a low entropy
component is separated from a large entropy component of the BEC fluid. This
would be similar to what happens in superfluid He-4.

In fact, one can think of two experimental set ups in order to check the
validity of the predictions made here. These experiments are analogous to the
ones done with superfluid He-4. Let us consider two containers ($A$ and $B$)
of a Bose-Einstein condensed fluid in which we keep the temperature initially
constant on each side and keep the density constant. At this point we consider
no external fields. Suppose that these containers are now physically connected
by a tube, in which we can have a flow of the condensate. Since
\begin{equation}
dg\,=\,\frac{1}{\rho}\,dP-s\,dT \tag*{(V.16)}%
\end{equation}
it follows from (V.14), that
\begin{equation}
\frac{1}{\rho}\,dP-s\,dT\,=\,-\,d\left(  \frac{m\,\vec{V}_{c}^{2}}{2}\right)
\ , \tag*{(V.17)}%
\end{equation}
so that the flow through the superleak occurs until $\,\vec{V}_{c}=0$. Under
this condition, it follows from (V.17) that
\begin{equation}
dP=s(T)\,dT\ . \tag*{(V.18)}%
\end{equation}

The temperature difference giving rise to a pressure difference is the
thermo\-mechanical effect. Equation~(V.18) is London relation \cite{3}.

If the containers discussed above are kept at constant pressure, then from
(V.16) it follows that
\begin{equation}
d\left(  \frac{m\,\vec{V}_{c}^{2}(x)}{2}\right)  =s\,dT\ . \tag*{(V.19)}%
\end{equation}

\vspace{0.2cm}

As a consequence of (V.19) one can see that if there is mass flow in one of
the containers, it will be colder than the other one. That is, the
mechanocaloric effect is expected to happen in any BEC system.

We can see, from (V.19), that in any BEC system a gradient temperature leads
to motion of the fluid and vice-versa. It follows, also from (V.19) that,
along a stream line, the difference between the kinetic energy of the fluid at
two points $B$ and $A$, in each container, is:
\begin{equation}
\frac{m\,\vec{V}_{c}^{2}(B)}{2}-\frac{m\,\vec{V}_{c}^{2}(A)}{2}%
\,=\,s(T)\left(  T_{B}-T_{A}\right)  \ . \tag*{(V.20)}%
\end{equation}

From equation (V.20) it follows that if the fluid flows from $\,A\,$ to
$\,B\,$ ($V(B)>V(A)$), the container $\,B\,$ will be warmer. On the other
hand, if one of the containers is now considered as a thick tube the pressure
$\,dP\,$ will cause a fountain of Bose-Einstein condensate. We predict, in
this way, the fountain effect for any BEC system.

In superfluid He-4 these two phenomena are explained by assuming that the
superfluid component is a zero-entropy fluid. We have seen here that this is
not necessary.

\bigskip

\section{\noindent\textbf{FORCE EQUATION} \vspace{0.3cm}}

In this section we will show that the phenomenon of superfluidity for a BEC
system is intimately connected to the two effects present in He-4, the
thermomechanic and mechanocaloric effects. Usually, they are considered as
independent properties of the superfluid Helium. We shall give, also in this
section, formal expressions for the thermoforce.

First of all we recall that
\begin{equation}
dg\,=\,\frac{1}{\rho}\,dP-s(T)\,dT\ . \tag*{(VI.1)}%
\end{equation}

Now we take the gradient of (V.14), and the dynamic equation for a nearly
uniform fluid becomes \vspace{0.2cm}
\begin{equation}
\frac{\partial\vec{P}_{c}}{\partial t}+m\left(  \vec{V}_{c}\cdot\vec{\nabla
}\right)  \vec{V}_{c}\,=\,\vec{F}_{\mathrm{ext}}\,-\,\frac{m}{\rho}%
\,\vec{\nabla}P+e\,\vec{V}_{c}\times\vec{B}+s(T)\,\vec{\nabla}T\ .
\tag*{(VI.2)}%
\end{equation}

\vspace{0.2cm}

All terms in equation~(VI.2) have a simple interpretation. The term in the
left hand side of equation (VI.2) is the total derivative of $\ \vec{P}_{c}$,
that is: \vspace{0.2cm}
\begin{equation}
\frac{D\vec{P}_{c}}{Dt}\;\equiv\;\left(  \frac{\partial}{\partial t}+\vec
{V}_{c}\cdot\vec{\nabla}\right)  \vec{P}_{c}\ . \tag*{(VI.3)}%
\end{equation}

\vspace{0.2cm}

The first term in the right hand side of (VI.2) is the external force
\begin{equation}
\vec{F}_{\mathrm{ext}}\,=\,-\vec{\nabla}U\ . \tag*{(VI.4)}%
\end{equation}

The second term in the right hand side of (VI.2) is the usual force that
induces motion in the fluid as a result of pressure gradients
\begin{equation}
\vec{F}_{1}\,=\,-\,\frac{m}{\rho}\,\vec{\nabla}P\ . \tag*{(VI.5)}%
\end{equation}

The third term in the r.h.s. of (VI.2) is the Lorentz force due to the
external magnetic field
\begin{equation}
\vec{F}_{\mathrm{Lorentz}}\,=\,e\,\vec{V}_{c}\times\vec{B}\ . \tag*{(VI.6)}%
\end{equation}

The last term in (VI.2) is our main contribution to the BEC motion. We propose
that this term is the thermoforce:
\begin{equation}
\vec{F}_{\mathrm{ther}}=s(T)\,\vec{\nabla}T\ . \tag*{(VI.7)}%
\end{equation}

The presence of the thermoforce in equation (VI.2) means that there will be
motion of the fluid as a result of temperatures gradients. The motion is
towards the greatest temperatures.

Putting all this together, we write
\begin{equation}
\frac{D\vec{P}_{c}}{Dt}\;\equiv\;\vec{F}_{\mathrm{ext}}+\vec{F}_{\mathrm{ther}%
}+\frac{1}{\rho}\,\vec{\nabla}P+e\,\vec{V}_{c}\times\vec{B}\ . \tag*{(VI.8)}%
\end{equation}

Except for the thermoforce, equation~(VI.8) is just the Euler equation for an
ideal fluid. Under similar assumptions made here, Anderson derived the usual
Euler equation for He-4 \cite{10}.

The conclusion, therefore, is that the same phenomenon\ (Bose-Einstein
condensation) is responsible for all the striking properties of superfluid
He-4: the superfluidity and the both related thermomechanical and
mechanocaloric effects. In fact, in principle, we expect that any BEC system
should present those phenomena. At least one of them, has already been
experimentally confirmed \cite{15}.

\section{\noindent\textbf{CONCLUSIONS} \vspace{0.3cm}}

We have shown in this paper that the thermomechanical and mechanocaloric
effects are always expected to occur in a Bose-Einstein condensate. Besides
the understanding of these effects, we have shown that they can be predicted
by using the same equation that leads to Euler equation (the equation that we
have named force equation).

We have proposed a generalized Bernoulli equation that would lead, for
non-uniform density fluids, to a departure from the usual Bernoulli equation.
The generalized Bernoulli equation is valid for stationary BEC states. For
time-independent states the chemical potential is zero, in this case the
Bose-Einstein condensed state is characterized as a zero Gibbs energy fluid.

Another important property of a superfluid is the fact that it
is irrotational \cite{Khalatnikov}. As a consequence of the presented
calculation, we have deduced that this is not generally the case. It only happens
in the absence of external magnetic fields.

We have derived an expression for the thermoforce. It is interesting to note
that the same expression was also found by London,\cite{3}, using a
different approach. Experimental evidence for the specific form (VI.7) is, to
our knowledge, missing. However, it predicts that as the temperature is
lowered the thermoforce tends to zero (since $\,\lim_{T\rightarrow
0}\limits\,s(T)\rightarrow0$). This fact is quite well known in superfluid
Helium-4. The thermoforce becomes negligible for temperatures below
$0.6^{\mathrm{o}}$K$\,$ \cite{3}.

Another important contribution of this present paper is that in order to
explain some aspects of the BEC phenomenon, there is no need of the zero
entropy fluid hypothesis.

Finally we would like to mention that some of our specific quantitative
predictions, contained in the presented equations can be tested using present
experimental conditions.

\bigskip

\section{\noindent\textbf{ACKNOWLEDGMENTS} \vspace{0.3cm}}

This work was partially supported by Funda\c{c}\~{a}o de Amparo \`{a} Pesquisa
do Estado de S\~{a}o Paulo (FAPESP), Conselho Nacional de Desenvolvimento
Cient\'{\i}fico e Tecnol\'{o}gico (CNPq) and by Programa de Apoio a
N\'{u}cleos de Excel\^{e}ncia (PRONEX).

\end{document}